\documentclass[conference]{IEEEtran}
\IEEEoverridecommandlockouts
\usepackage{cite}
\usepackage{amsmath,amssymb,amsfonts}
\usepackage{algorithmic}
\usepackage{graphicx}
\usepackage{textcomp}
\usepackage{xcolor}
\def\BibTeX{{\rm B\kern-.05em{\sc i\kern-.025em b}\kern-.08em
    T\kern-.1667em\lower.7ex\hbox{E}\kern-.125emX}}
\begin{document}

\title{Federated Learning under Attack: Improving Gradient Inversion for Batch of Images\\
}

\author{\IEEEauthorblockN{Luiz~Leite\IEEEauthorrefmark{1}, Yuri~Santo\IEEEauthorrefmark{1}, Bruno~L.~Dalmazo\IEEEauthorrefmark{2}, André~Riker\IEEEauthorrefmark{1}}
\IEEEauthorblockA{\IEEEauthorrefmark{1}Federal University of Par\'a, Brazil}
\IEEEauthorblockA{\IEEEauthorrefmark{2} Federal University of Rio Grande, Brazil}%
\\[-3.0ex]
}

\maketitle

\begin{abstract}

Federated Learning (FL) has emerged as a machine learning approach able to preserve the privacy of user's data. Applying FL, clients train machine learning models on a local dataset and a central server aggregates the learned parameters coming from the clients, training a global machine learning model without sharing user's data. However, the state-of-the-art shows several approaches to promote attacks on FL systems. For instance, inverting or leaking gradient attacks can find, with high precision, the local dataset used during the training phase of the FL. This paper presents an approach, called Deep Leakage from Gradients with Feedback Blending (DLG-FB), which is able to improve the inverting gradient attack, considering the spatial correlation that typically exists in batches of images. The performed evaluation shows an improvement of 19.18\% and 48,82\% in terms of attack success rate and the number of iterations per attacked image, respectively.

\end{abstract}


\section{Introduction}
Privacy-preserving solutions are a key requirement for almost all computer applications, whether for legislation compliance or due to the mistrust of how sensitive data can be used. In an era marked by escalating concerns over data breaches and privacy violations, ensuring the confidentiality and integrity of personal information has become paramount for businesses and individuals alike \cite{mothukuri2021survey}. Moreover, as technology continues to advance, the need for robust privacy-preserving techniques becomes even more pressing. 


 Federated Learning (FL) emerges as a solution to provide data-privacy for smart systems because it enables distributed Machine Learning (ML) training, without sending user's data to a central point, providing an extra level of user data-privacy protection \cite{10180041}. In FL, multiple ML models run on local privacy-sensitive datasets, simultaneously, and a global ML model, running on a server is built without sharing the local datasets with the server. 

Some FL methods rely on gradient or weight sharing between clients and servers to train the global ML model \cite{geiping2020inverting}. Historically, there was a prevalent belief that sharing gradients was inherently secure, implying that the exchange does not compromise the confidentiality of the training data. However, the authors of \cite{zhu2019deep} proposed a method called Deep Leakage from Gradients (DLG), showing how to invert the gradient to reconstruct the input data used in the training, becoming one of the most famous methods to attack FL models. 

This paper aims to enhance the performance of DLG-based methods, proposing a new attack algorithm called  Deep Leakage from Gradients with Feedback Blending (DLG-FB). The proposed approach takes advantage of the spatial redundancies in a batch of images. Assuming the attacker aims to access the whole batch of images in a local dataset, DLG-FB does not initialize the input image-matrix with dummy data, i.e. pure random values. Instead, DLG-FB computes a blend of images that have already been accessed by the attacker after the sequence of attack is greater than two reconstructed images. The performance evaluation shows that DLG-FB improves the capacity to attack images and reduces the number of iterations to reach a successful attack.

This paper is organized as follows: 
Section \ref{sec:RW} provides a review of the main related works. Section \ref{sec:threat-model} presents the threat model considered in this work.
Section \ref{sec:solution} describes the proposed attack. Section \ref{sec:results} presents the conducted evaluation and the obtained results. Section \ref{sec:conclusion} introduces the conclusions and outlines potential directions for future research.

\section{Related Work}
\label{sec:RW}

Gradient exchange is one of the prevalent techniques in contemporary multi-node machine learning setups, such as distributed training and collaborative learning, as Federated Learning. This section aims to describe the most relevant attacks able to exploit gradient in FL systems.

Deep Leakage from Gradients (DLG) \cite{zhu2019deep} and Improved Deep Leakage from Gradients (iDLG) \cite{zhao2020idlg} show how it is possible to leak private training data if an attacker has access to the shared gradients.

\subsection{Deep Leakage from Gradients (DLG)}

The work proposed by \cite{zhu2019deep} presents an iterative method, called Deep Leakage from Gradients (DLG),  based on an optimization algorithm that can obtain both the training inputs and the labels, considering an attacker accessing the gradient coming from a FL client, defined as \textit{$\nabla$W}.

\begin{figure}
    \centering
    \includegraphics[width=1\linewidth]{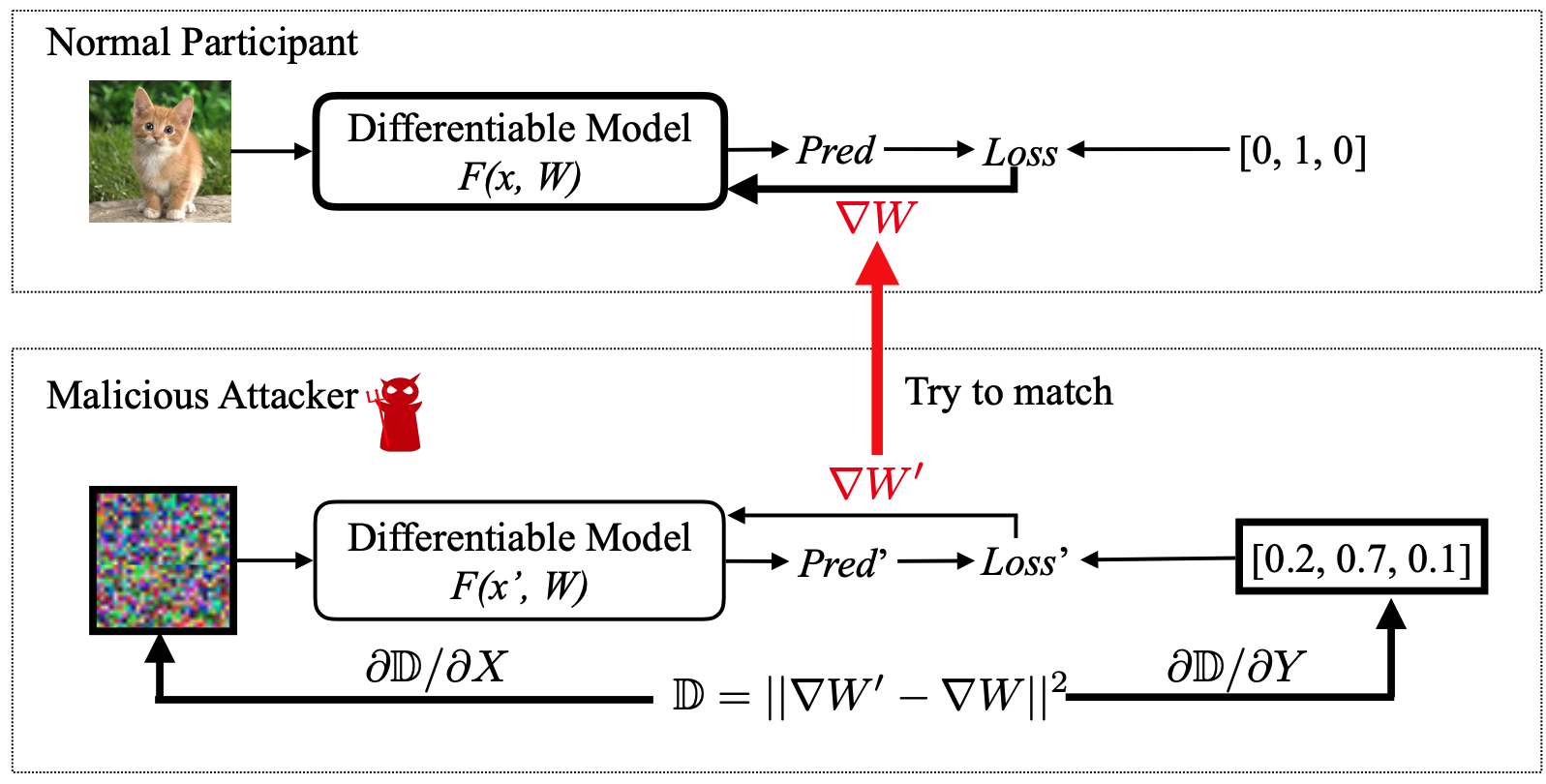}
    \caption{DLG Algorithm \cite{zhu2019deep}.}
    \label{fig:DLG}
\end{figure}

The first step of DLG, after accessing the client gradient, is to randomly initialize a dummy input and label input. With the “dummy data” it is possible to compute “dummy gradients”, defined as \textit{$\nabla$W'}. As depicted in Fig. \ref{fig:DLG}, this attack seeks to approximate \textit{$\nabla$W'} to \textit{$\nabla$W}, changing iteratively the input and the label data. When the \textit{$\nabla$W'} is close to \textit{$\nabla$W}, it is possible to extract the data used by the client to train the machine learning model.

As can be noticed, this is an optimization problem, where the distance $||\nabla W' - \nabla W||^2$ is differentiable concerning inputs and labels. This optimization is computed by the solver called Limited-memory Broyden, Fletcher, Goldfarb, and Shanno (L-BFGS).

\subsection{Improved Deep Leakage from Gradients (iDLG)}
The work proposed by \cite{zhao2020idlg} aims to improve the DLG method. In DLG, the attacker generates dummy data and corresponding labels under the guidance of shared gradients. Authors in \cite{zhao2020idlg} proposed Improved Deep Leakage from Gradients (iDLG), which exploits the relationship between the ground-truth labels and the signs of the gradients.  This work shows that label information can be computed analytically from the gradients, and this information can be used to obtain train data more close to the original. As another advantage, this method is suitable for any differentiable model trained with cross-entropy loss on one-hot labels.

\subsection{Other Relevant Attacks}
Recovery of image data from gradient information was first discussed in \cite{phong2017privacy} for neural networks. In this work, authors have proven that recovery is possible for a single neuron or linear layer.

In \cite{sannai2018reconstruction}, the authors also address the leakage problem in deep learning. This work proposes a method to obtain the sample trained data for deep-learning models based on the ReLu function. However, for this method, it is necessary to access the entire learning process.

In another effort, the authors discuss the recovery of training data from shared gradients in distributed machine learning systems~\cite{10.1155/2023/5510329}. The original Deep Leakage from Gradients (DLG) method faces issues with accuracy and stability due to exploding gradients and high learning rates. To address these issues, this study proposes the WDLG method, which uses the Wasserstein distance to calculate loss, enabling more faithful and efficient recovery of training data. 


\section{Threat Model}
\label{sec:threat-model}
For the proposed approach in this work, we consider the scenario of an honest but curious federated learning server seeking to access user data, i.e. a batch of images. The batch of image, which is the target of the attacker, is defined as $S=\{s_1, s_2, ..., s_k\}$, where $s_k$ is the last image to be reconstructed. The attacker has access to the same machine learning model architecture, as it is expected for a federated learning server. Besides, the attacker can store and process updates sent by individual users independently but lacks the ability to interfere with the collaborative learning algorithm. Moreover, the attacker is unable to alter the model architecture to facilitate their attack or send deceptive global parameters that do not accurately represent the learned global model.

\section{Deep Leakage from Gradients with Feedback Blending (DLG-FB)}
\label{sec:solution}

In prior gradient-based attack methodologies, random noise, also known as dummy data, has traditionally been used to initialize the data input fed into the first iteration of the Limited-memory Broyden-Fletcher-Goldfarb-Shanno (L-BFGS) solver. However, with random data initialization, when attempting to reconstruct two consecutive attacked images, i.e., $s_n$ and $s_{n+1}$, traditional approaches start with a new random guess for each iteration, meaning that each image reconstruction is entirely independent of the others.

Consequently, although it is possible to gather information from the successful reconstructed images during the attacked sequence, this knowledge is not effectively used by the attacker.

\begin{figure*}
    \centering
    \includegraphics[width=0.7\linewidth]{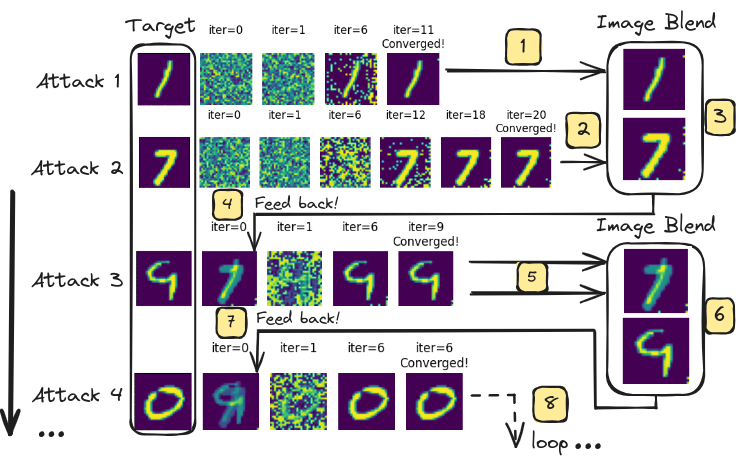}
    \caption{DLG-FB algorithm illustration.}
    \label{fig:fb_ilustration}
\end{figure*}

In response to this scenario, we introduce a novel DLG-based attack strategy defined as Deep Leakage from Gradients with Feedback Blending (DLG-FB), as illustrated in Figure~\ref{fig:fb_ilustration}. This approach is designed for attacks that target a batch of images, which means the attacker aims to reconstruct a set of images from the same federated learning client. We assume these images have some level of spatial redundancy, keeping a certain degree of similarity. For instance, it is expected the attacked client has a batch of images with the same set of people or place. During the sequence to attack a batch of images, the main idea of DLG-FB is to blend successful reconstructed images and feed the solver with the blended image, making the blended image a better starting point for the attack than pure random data.

DLG-FB algorithm requires two successfully obtained images, as can be observed in labels 1 and 2 in Figure~\ref{fig:fb_ilustration}. After two reconstructed images, DLG-FB performs the images blending (see label 3) from attack 1 and 2 and uses it as initial guess for the attack 3 (label 4). This process keeps repeating for the next elements of the attacked batch. However, it is important to notice that the reconstruction can fail, due to a vast number of factors.

If an image is successfully reconstructed, it blends the new image with the previous blend to create a new composite, as can be seen in labels 5, 6 and 7 of Figure~\ref{fig:fb_ilustration}. In case of failure, it proceeds to the next image without incorporating the failed reconstructed image, which would introduce excessive noise. 

Regarding the image blending performed by DLG-FB, $C$ represents the pixel matrix of an image, where $C_o$ is the resulting composite image, $C_a$ and $C_b$ are the two reconstructed images. In this work, the blending is determined by the following equation:

\begin{equation}
    C_o = \alpha C_a + (1 - \alpha) C_b
    \label{eq:alpha}
\end{equation}

In equation \ref{eq:alpha}, $\alpha$ is an arbitrarily chosen value that determines which image will be more dominant in the composition.


\section{Evaluation and Results}
\label{sec:results}

This section presents the evaluation of the proposed attack algorithm. First, the test environment and the dataset used are introduced. Then, the obtained results are discussed.

\subsection{Test Environment and Dataset}
In our testing environment, we employed the PyTorch API within a Python virtual environment running on Manjaro Linux x86 64 OS, powered by an AMD Ryzen 5 5600G CPU, 4.464GHz. Besides, we utilized PyTorch's LBFGS (Limited-memory Broyden-Fletcher-Goldfarb-Shanno) optimizer with a learning rate set to 1. 

For the performance evaluation of the approaches, we have used two well-known datasets: CIFAR100 and MNIST.

\begin{itemize}
    \item CIFAR-100 is a widely used computer vision dataset comprising 60,000 color images grouped into 100 classes, each with 600 images. These 32x32 pixel images span diverse objects, animals, and scenes, organized into 20 superclasses. 

    \item MNIST is a well-known dataset utilized in computer vision which comprises 70,000 grayscale images of handwritten digits, categorized into 10 classes. MNIST serves as a foundational benchmark for machine learning models.
\end{itemize}


This optimization algorithm is well-suited for large-scale problems, particularly in machine learning. Unlike the full BFGS algorithm, LBFGS conserves memory by storing only a few vectors to approximate the Hessian matrix, exploiting curvature information for faster convergence. It belongs to the family of quasi-Newton methods, efficiently approximating second-order derivatives using first-order information.




\subsection{Obtained Results}
We have compared the proposed DLG-FB with the following approaches: (i) Original DLG, (ii) original iDLG, (iii) iDLG-FB (iv) DLG-FB-Noise Factor (iDLG-FB-NF), and (v) DLG-FB-Noise Factor (DLG-FB-NF).


It is important to note that iDLG-FB represents the implementation of the proposed Feedback Blending strategy in iDLG. Additionally, for comparison purposes, DLG-FB-NF and iDLG-FB-NF are versions of the proposed Feedback Blending strategy that do not discard the dummy data from unsuccessful attack attempts. Instead, they blend it in, aiming to reduce overfitting. These versions always blend the output image, even if the attack sequence fails. Moreover, for the obtained results shown in this section, $\alpha = 0.5$ (see eq. \ref{eq:alpha}).

\begin{figure}
\centering     
\includegraphics[width=0.9\linewidth]{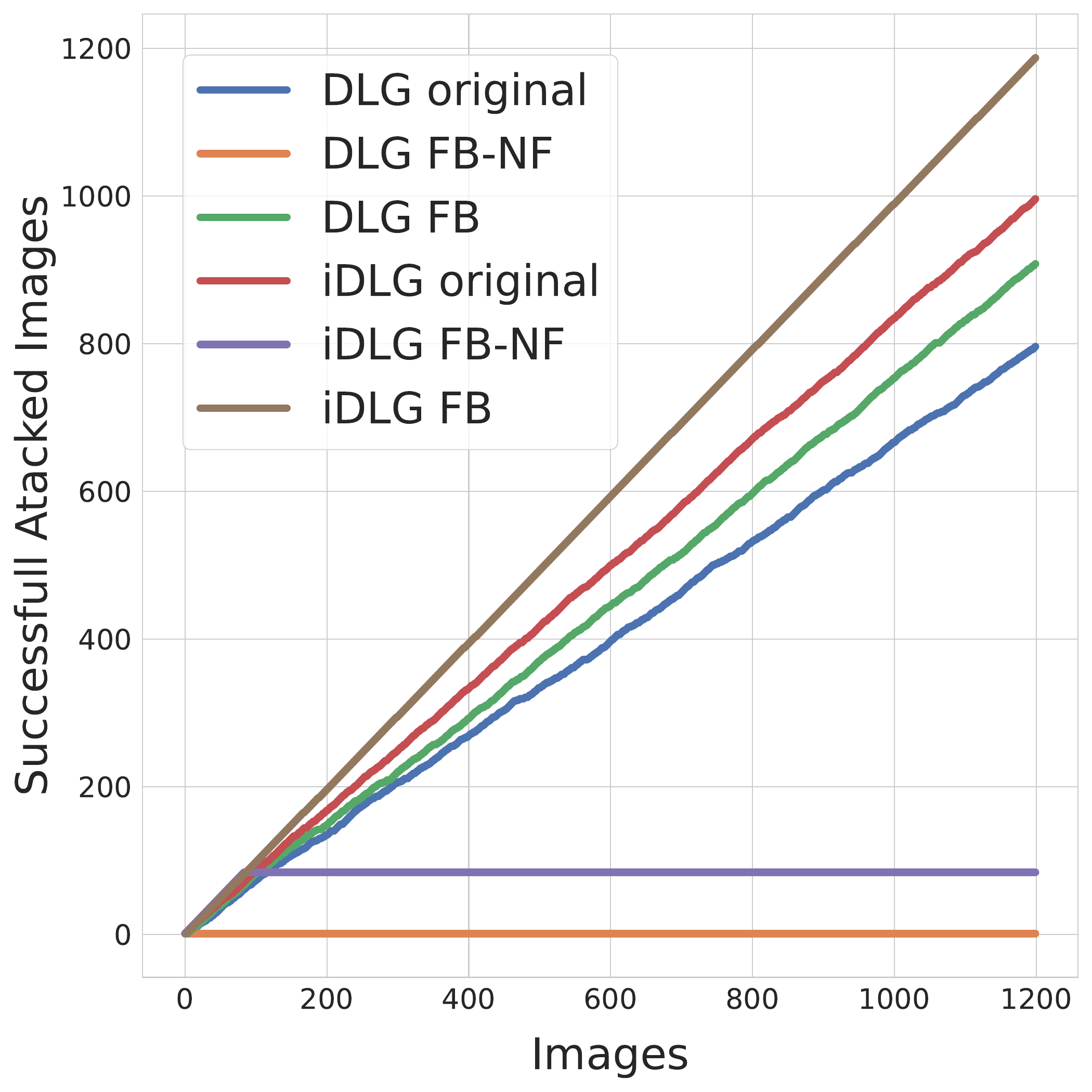}

\caption{Cumulative Number of Successful Reconstructed Image (CIFAR-100).}
\label{fig:convergencia1}
\end{figure}

\begin{figure}
\centering     

\includegraphics[width=0.9\linewidth]{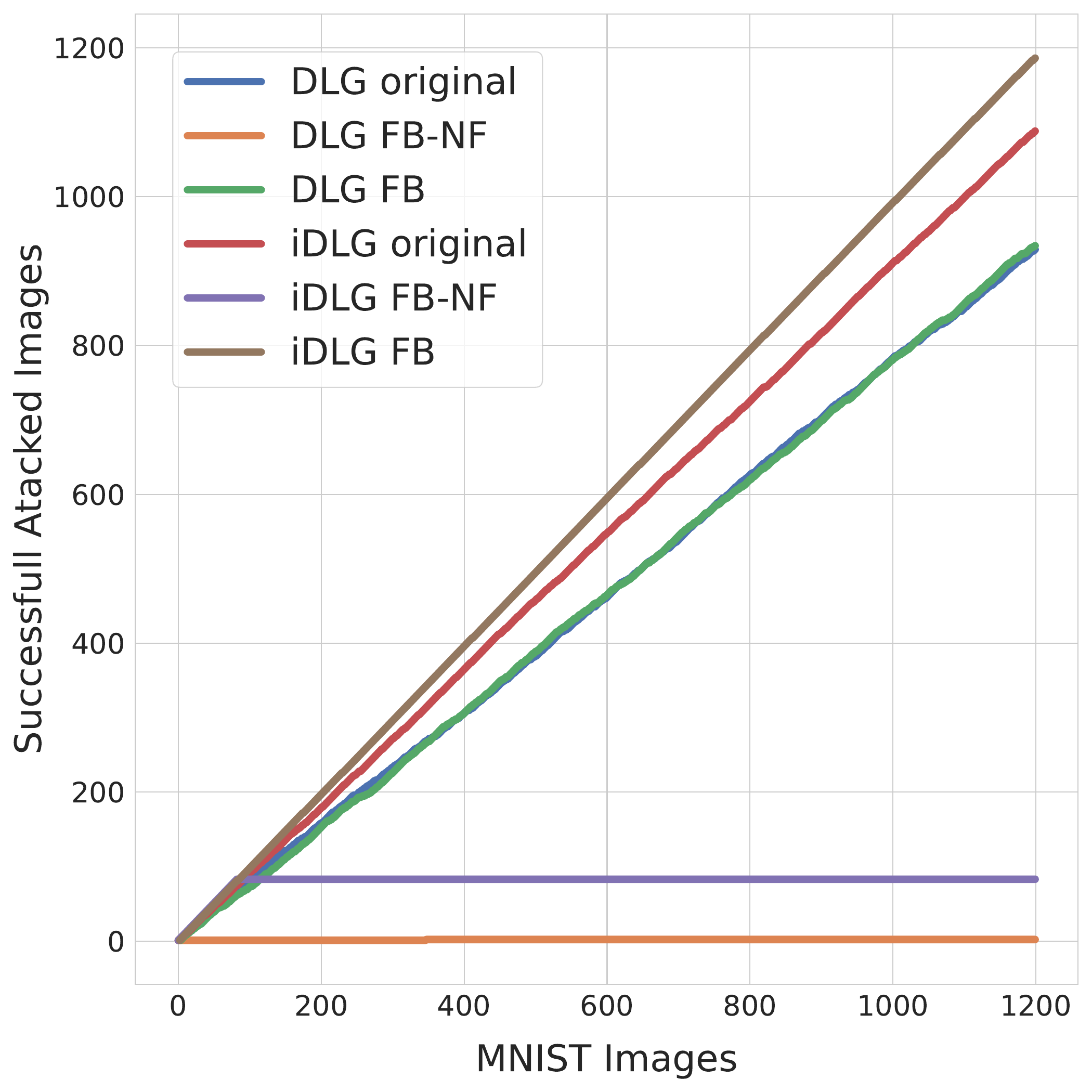}
\caption{Cumulative Number of Successful Reconstructed Image (MNIST).}
\label{fig:convergencia2}
\end{figure}

Figure \ref{fig:convergencia1} and  \ref{fig:convergencia2} present the obtained results in terms of the number of successfully attacked images for the CIFAR-100 and MNIST datasets, respectively. For CIFAR-100, the original iDLG recovered 996 images, while iDLG-FB recovered 1187 images, a +19.18\% improvement. For DLG-based approaches, original DLG was able to recover 769 images compared to 908 images by the DLG-FB, a +14.07\% increase. Regarding MNIST, which consists of simpler grayscale images with no color channels, original iDLG recovered 1088 images, while iDLG-FB recovered 1186 images, a +9.01\% improvement. For MNIST, original DLG recovered 929 images compared to 934 by the DLG-FB version. The FB versions assist the solver by providing better initial guesses for the pixels. However, since grayscale images have significantly fewer pixels, the new strategy has a reduced impact.

\begin{figure}
\centering     
\includegraphics[width=0.9\linewidth]{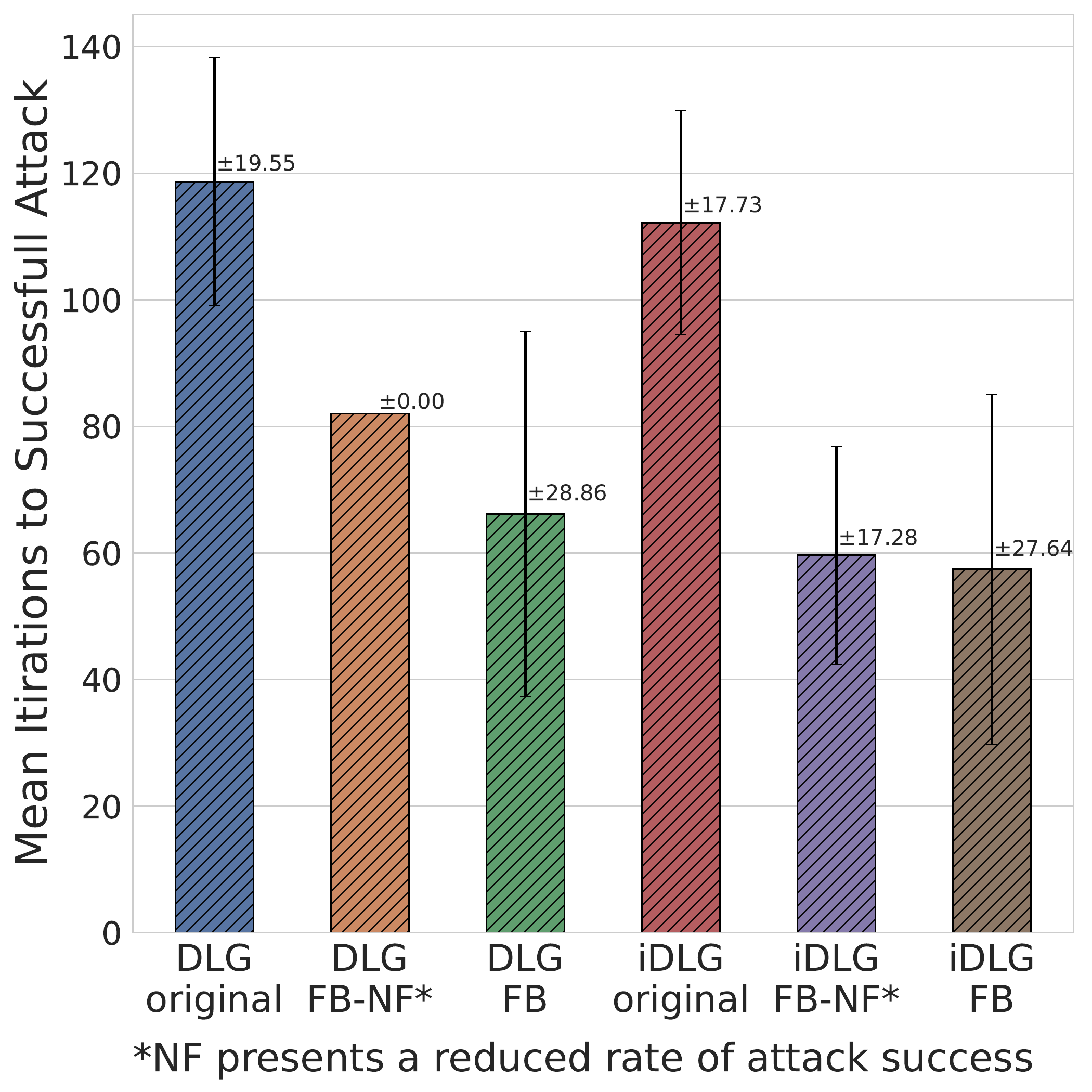}
\caption{Mean Number of Iterations to Successful Image Reconstruction.}
\label{fig:iteracoes1}
\end{figure}

\begin{figure}
\centering     
\includegraphics[width=0.9\linewidth]{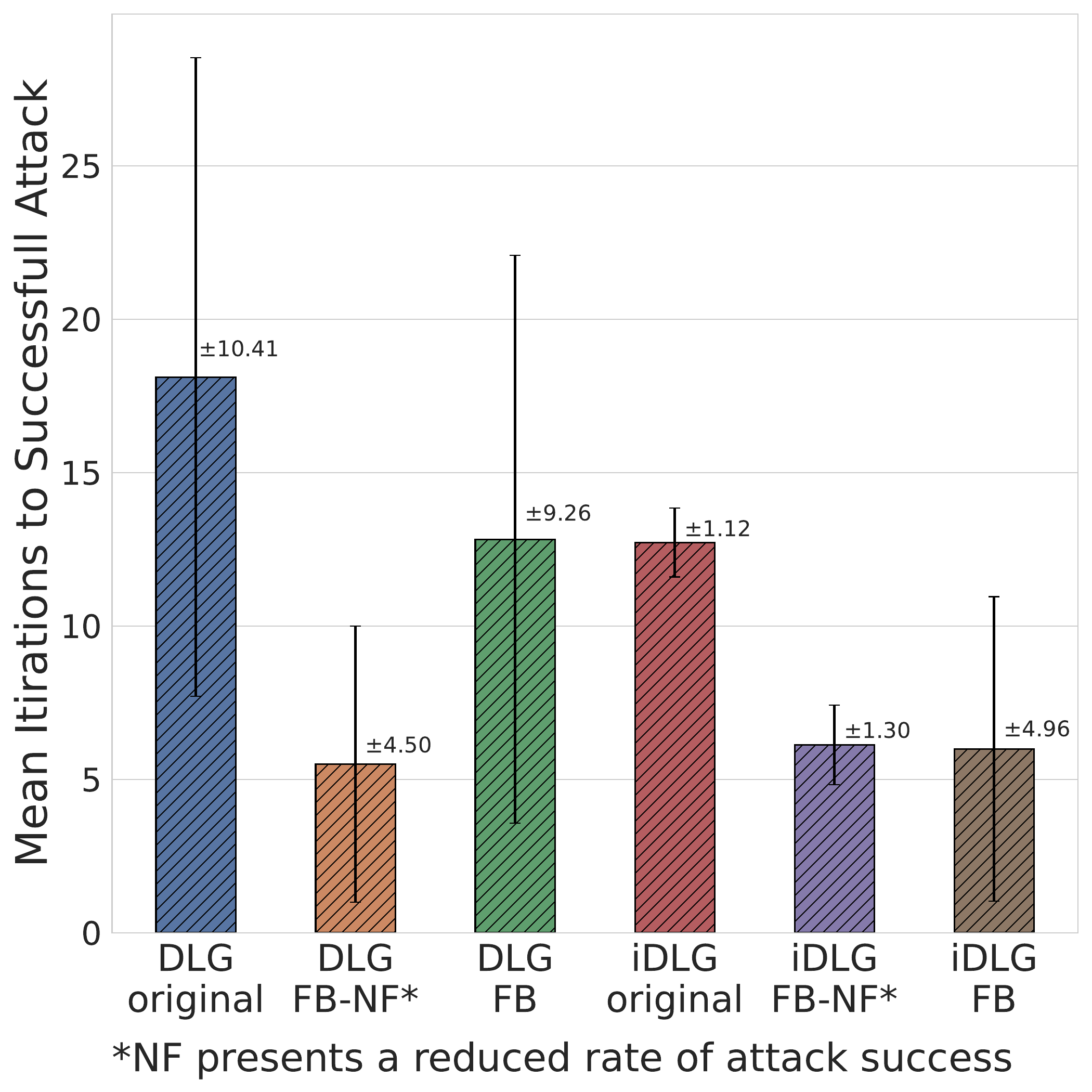}
\caption{Mean Number of Iterations to Successful Image Reconstruction.}
\label{fig:iteracoes2}
\end{figure}

As depicted in Figure\ref{fig:convergencia1} and \ref{fig:convergencia2}, the two approaches using the Feedback Blending Noise Factor (FB-NF) demonstrate a very peculiar behavior: at a certain point when noise becomes too prevalent, it fails to reconstruct the targeted image, since it is designed to not take into consideration if the output image has converged to the attacked image.
Figure \ref{fig:iteracoes1} and \ref{fig:iteracoes2} present the mean number of iterations to reach a successful image attack for CIFAR100 and MNIST datasets, respectively. In CIFAR100 dataset, DLG-FB and iDLG-FB  had 48,82\% and 44,26\% less iterations than original DLG and iDLG, respectively. In MNIST, the proposed Feedback Blending strategy has shown 52.94\% and 29,19\% less iterations when applied to iDLG and DLG, respectively.

It is worth to mention that LBFGS faces greater challenges with CIFAR-100 than MNIST due to its more complexity images and higher pixel count from multiple color channels. MNIST is single grayscale channel and lower pixel count make it more susceptible to such attacks.

\begin{figure*}
    \centering
    \includegraphics[width=1\linewidth]{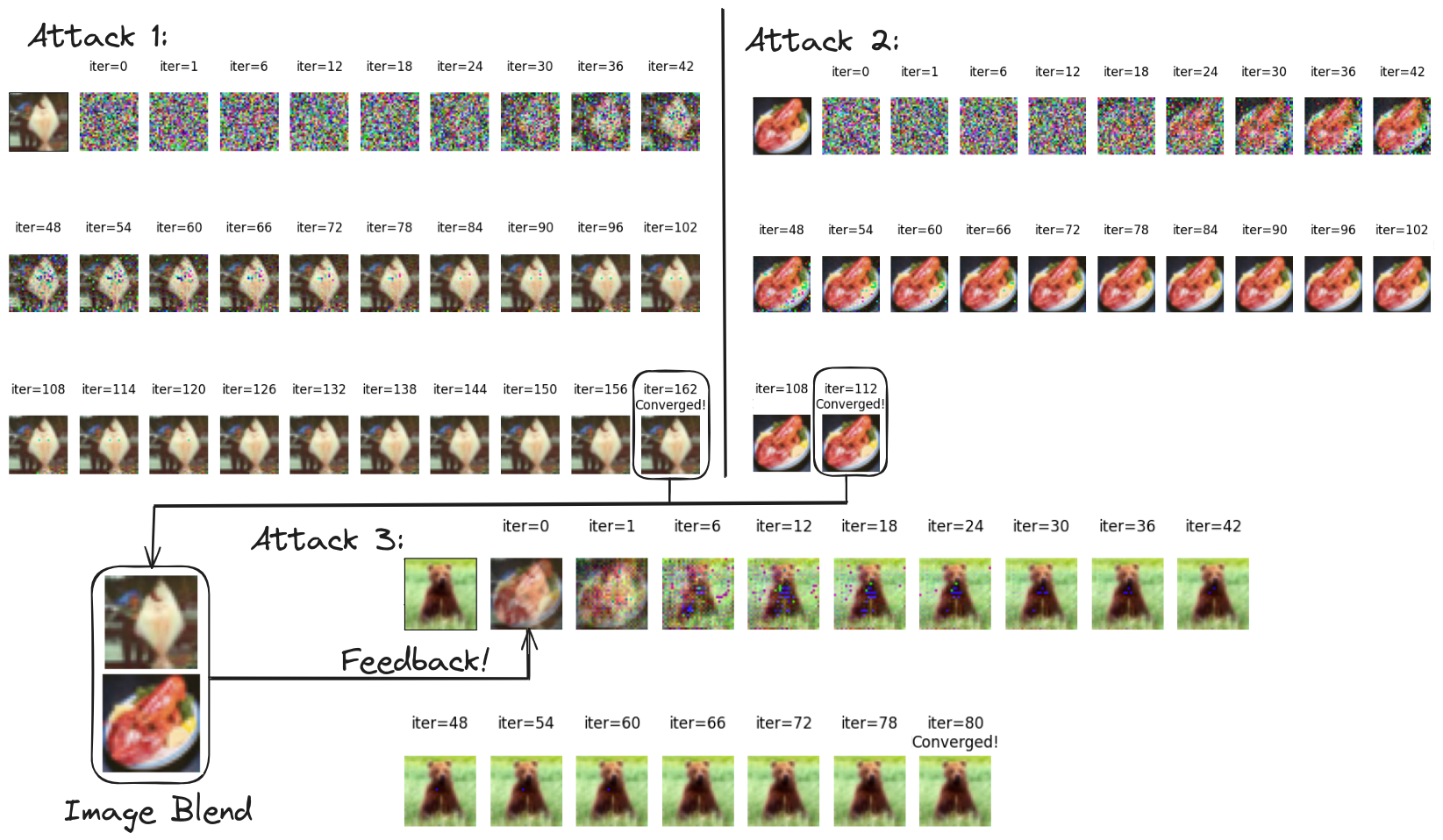}
    \caption{Samples of DLG-FB Reconstructing Images (CIFAR100).}
    \label{fig:fb_cifar100_ilustration}
\end{figure*}

Fig. \ref{fig:fb_cifar100_ilustration} shows a sequence of iterations and the feedback blending performed by DLG-FB for three sample images in CIFAR100 dataset.

\section{Conclusion and Future Works}
\label{sec:conclusion}
Privacy-preserving solutions have become a critical necessity in almost all computer applications. Among the possible solutions to this issue, Federated Learning (FL) has emerged as a promising approach for safeguarding data privacy in smart systems. In this context, this paper introduces a novel approach, named Deep Leakage from Gradients with Feedback Blending (DLG-FB), aimed at enhancing the effectiveness of DLG-based methods in federated learning systems. DLG-FB takes advantage on the spatial redundancies present in batches of images. Unlike conventional methods that initialize input image-matrices with random data to attack a single image, DLG-FB employs a strategy to attack a entire batch of image. After more than two successful image reconstructions, DLG-FB computes a blend of images and uses it as the initial data rather than utilizing pure random values. DLG-FB reveals gains in both the number of images successfully attacked and the iterations required to achieve a successful attack.



As future work, the authors intend to add more mechanisms to the proposed DLG-FB, improving the criteria to perform the image blend. Also, a machine learning model can be used to make the approach more efficient. Moreover, the new version of the attack will be tested using other image datasets.

\section*{Acknowledgements}

This study was financed in part by: (i) FAPERGS/CNPq (23/2551-0000773-8), (ii) UE iTec/FURG (iTec-80), (iii) PROAP/UFPA.

\bibliographystyle{ieeetr}
\bibliography{paper}

\end{document}